%% file: MGarXiv.tex
\documentclass[aps,prl,twocolumn,showpacs,floatfix,longbibliography]{revtex4-1}
\usepackage{graphicx,amsfonts,amssymb,amsmath}
\usepackage{textcomp}
\usepackage{esint}
\usepackage[colorlinks,linktocpage,bookmarks=false,citecolor=blue,linkcolor=red,urlcolor=blue]{hyperref}
\usepackage[T1]{fontenc}
\usepackage[dvipsnames]{xcolor}
\usepackage[english]{babel}
\usepackage{afterpage}

\allowdisplaybreaks

\def\vLL{v_{\rm LL}}
\def\KLL{K_{\rm LL}}

\begin{document}

\input{./definition.tex}

\graphicspath{{./figures_submit/}}

\title{Mott-glass phase of a one-dimensional quantum fluid with long-range interactions} 
\author{Romain Daviet and Nicolas Dupuis}
\affiliation{Sorbonne Universit\'e, CNRS, Laboratoire de Physique Th\'eorique de la Mati\`ere Condens\'ee, LPTMC, F-75005 Paris, France}

\date{June 1, 2023} 

\begin{abstract}
We investigate the ground-state properties of quantum particles interacting {\it via} a long-range repulsive potential ${\cal V}_\sigma(x)\sim 1/|x|^{1+\sigma}$ ($-1<\sigma$) or ${\cal V}_\sigma(x)\sim -|x|^{-1-\sigma}$ ($-2\leq \sigma <-1$) that interpolates between the Coulomb potential ${\cal V}_0(x)$ and the linearly confining potential ${\cal V}_{-2}(x)$ of the Schwinger model. In the absence of disorder the ground state is a Wigner crystal when $\sigma\leq 0$. Using bosonization and the nonperturbative functional renormalization group we show that any amount of disorder suppresses the Wigner crystallization when $-3/2<\sigma\leq 0$; the ground state is then a Mott glass, i.e., a state that has a vanishing compressibility and a gapless optical conductivity. For $\sigma<-3/2$ the ground state remains a Wigner crystal.   	
\end{abstract}

\maketitle

\paragraph{Introduction.} 

The ground state of a one-dimensional quantum fluid with short-range interactions is generically a Luttinger liquid. This corresponds to a metallic state, which is, however, not described by Landau's Fermi liquid theory, for fermions and to a superfluid state, but without Bose-Einstein condensation, for bosons~\cite{Giamarchi_book}. In the presence of disorder, the ground state either remains a Luttinger liquid or becomes an Anderson insulator (fermions) or a Bose glass (bosons), i.e., an insulating state with a vanishing dc conductivity, a gapless optical conductivity and a nonzero compressibility~\cite{Giamarchi87,Giamarchi88,Fisher89}. 

Whether one-dimensional disordered quantum fluids can exhibit other phases besides the Luttinger liquid and the Anderson-insulator or Bose-glass phases has been the subject of debate for a long time. In particular, several works have addressed the existence of a Mott-glass phase but no firm positive conclusion has been reached so far. The Mott glass is intermediate between the Mott insulator and the Anderson insulator or Bose glass, and is characterized by a vanishing compressibility and a gapless conductivity; it would result from the coexistence of gapped single-particle excitations (which imply a vanishing compressibilty) and gapless particle-hole excitations (hence the absence of gap in the conductivity). 

On the one hand it has been proposed that the interplay between disorder and a commensurate periodic potential could stabilize a Mott glass~\cite{Orignac99,Giamarchi01} but this conclusion, when the interactions are short range, has been challenged~\cite{Nattermann07,Ledoussal08a}. On the other hand, the existence of a Mott glass in a disordered system with linearly confining interactions mediated by a $(1+1)$-dimensional gauge field (disordered Schwinger model) has been predicted by the Gaussian variational method~\cite{Chou18} and the perturbative functional renormalization group (FRG)~\cite{Giamarchi01}, but this conclusion is in conflict with a recent study based on the nonperturbative FRG~\cite{Dupuis20a}. The only system that seems to certainly satisfy the basic properties of the Mott glass is the one-dimensional electron gas with (unscreened) Coulomb interactions~\cite{Shklovskii81}.

\begin{figure}[b]
	\centerline{\includegraphics[width=7.5cm]{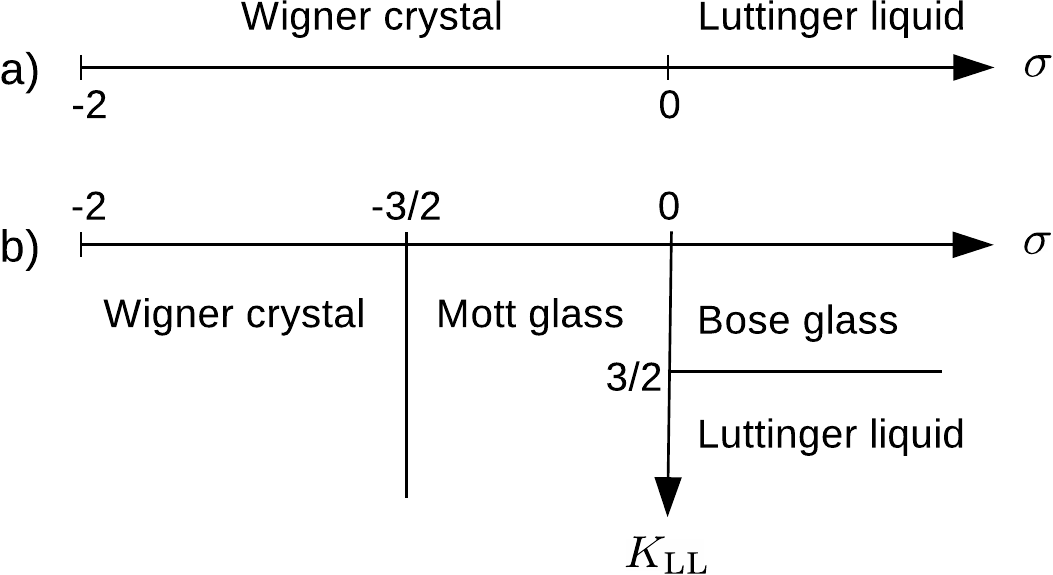}}
	\caption{Phase diagram of a pure (a) or disordered (b) one-dimensional Bose fluid with short-range interactions and a long-range interaction potential $\calV_\sig(x)$ [Eq.~(\ref{Vsig})]. $\KLL$ denotes the Luttinger parameter associated with short-range interactions; it determines the ground state when $\sig>0$ (in that case $\calV_\sig$ is effectively short range) but plays no important role for long-range interactions.}
	\label{fig_phase_dia} 
\end{figure} 

In this Letter we determine the phase diagram of a one-dimensional quantum fluid where the particles interact with both a short-range potential and a long-range potential
\beq 
\calV_\sig(x) = \left\{ 
\begin{array}{lll}
	\frac{e^2}{(x^2+a^2)^{(1+\sig)/2}} & \mbox{if} & -1 < \sig , \\ 
	- e^2 \ln|x/a| & \mbox{if} & \sig=-1 , \\
	- e^2 |x|^{-1-\sig} &  \mbox{if} & -2\leq \sig <-1 
\end{array}
\right.
\label{Vsig} 
\eeq
($a$ is a short-distance cutoff~\cite{not4}) that interpolates between the Coulomb potential $\calV_0(x)$ and the linearly confining potential $\calV_{-2}(x)$ of the Schwinger model~\cite{Schwinger62,Coleman76}. Although our conclusions hold for both fermions and bosons, we use the terminology of the Bose fluid in the following. 

Our main results are summarized in Fig.~\ref{fig_phase_dia}. The ground state of the pure fluid is a Luttinger liquid for $\sig>0$ (in that case $\calV_\sig$ is effectively short range) and a Wigner crystal for $\sig\leq 0$ as first shown by Schulz~\cite{Schulz93,Fano99,Fogler05,Casula06,Astrakharchik11} in the case of Coulomb interactions (true long-range crystalline order however occurs only for $\sig<0$). In the presence of disorder, the Wigner crystal is stable if the interactions are sufficiently long range, i.e., $\sig<-3/2$, but is unstable against a Mott glass when $-3/2<\sig\leq 0$. Apart from the vanishing compressibility, we find that the Mott glass is described by a fixed point of the FRG flow equations similar to the one describing the Bose-glass phase. Besides the finite localization length and the gapless conductivity, this fixed point is characterized by a renormalized disorder correlator that assumes a cuspy functional form whose origin lies in the existence of metastable states associated with glassy properties~\cite{Dupuis19,Dupuis20}.

\paragraph{Model and FRG formalism.} 

The low-energy Hamiltonian of the pure Bose fluid in the presence of the long-range interaction potential $\calV_\sig(x)$ can be written as 
\beq
\hat H = \HLL + \half \sum_q \hat\rho(-q) \calV_\sig(q) \hat\rho(q) , 
\eeq
where $\hat\rho(q)$ and $\calV_\sig(q)$ are the Fourier transforms of $\hat\rho(x)$ and $\calV_\sig(x)$, and a UV momentum cutoff $\Lambda$ is implied. The Luttinger-liquid Hamiltonian $\HLL$ includes the kinetic energy of the particles and their short-range interactions. In the bosonization formalism~\cite{Giamarchi_book}, 
\begin{align}
\hat H ={}&  \sum_q \frac{\vLL q^2}{2\pi} \llbrace \frac{1}{\KLL} \hat\varphi(-q)\hat\varphi(q) + \KLL \hat\theta(-q) \hat\theta(q) \rrbrace 
\nonumber \\ & + \frac{1}{2\pi^2} \sum_q q^2 \calV_\sig(q) \hat\varphi(-q) \hat\varphi(q) , 
\label{ham1}
\end{align}
where $\hat\theta$ is the phase of the boson operator $\hat\psi(x)=e^{i\hat\theta(x)}\hat\rho(x)^{1/2}$. $\hat\varphi$ is related to the density operator {\it via} 
\beq
\hat\rho(x) = \rho_0 - \frac{\dx\hat\varphi(x)}{\pi} + 2 \sum_{m=1}^\infty \rho_{2m} \cos(2m\pi\rho_0x-2m\hat\varphi(x)) ,
\eeq 
where $\rho_0$ is the average density and the $\rho_{2m}$'s are nonuniversal parameters that depend on microscopic details. $\hat\varphi$ and $\hat\theta$ satisfy the commutation relations $[\hat\theta(x),\partial_y\hat\varphi(y)]=i\pi\delta(x-y)$. $\vLL$ denotes the velocity of the sound mode when $\calV_\sig=0$ and the dimensionless parameter $\KLL$, which encodes the strength of the short-range interactions, is the Luttinger parameter. 

In the absence of long-range interactions ($\calV_\sig(q)=0$), the system is a Luttinger liquid, characterized by a nonzero compressibility $\kappa=\KLL/\pi\vLL$ and a nonzero charge stiffness or Drude weight (defined as the Dirac peak $\delta(\w)$ in the conductivity) $D=\vLL\KLL$~\cite{not3}. The superfluid correlation function $\mean{\hat\psi(x)\hat\psi^\dagger(0)}\sim 1/|x|^{1/2\KLL}$ and the density correlation function $\mean{\hat\rho(x)\hat\rho(0)}_{|q|\sim 2\pi\rho_0}\sim \cos(2\pi\rho_0x)/|x|^{2\KLL}$ decay algebraically; the former dominates for $\KLL>1/2$, the latter for $\KLL<1/2$ (all other correlation functions are subleading). 

The long-range interaction potential $\calV_\sig(q)$ can be simply taken into account by introducing momentum-dependent velocity and Luttinger parameter defined by 
\beq 
\begin{gathered} 
v(q)K(q) = \vLL\KLL , \\ 
\frac{v(q)}{K(q)} = \frac{\vLL}{\KLL} + \frac{\calV_\sig(q)}{\pi} . 
\end{gathered}
\label{Kqvq} 
\eeq 
The long-range potential in~(\ref{ham1}) can then be simply taken into account by replacing, in the Luttinger-liquid Hamiltonian, $\vLL$ and $\KLL$ by $v(q)$ and $K(q)$~\cite{Giamarchi_book}. For $\sig>0$, since $\calV_\sig(q)$ has a finite limit for $q\to 0$, $v(q=0)$ and $K(q=0)$ are finite; this essentially leads to a mere renormalization of $\vLL$ and $\KLL$ and the ground state remains a Luttinger liquid. By contrast, for $\sig\leq 0$, in the small-momentum limit $\calV_\sig(q)\sim |q|^\sig$ so that $v(q)\sim |q|^{\sig/2}$ and $K(q)\sim |q|^{-\sig/2}$ are determined by the long-range part of the interactions (for $\sig=0$, $|q|^\sig$ should be interpreted as $-\ln|q|$), which drastically modifies the ground state and the low-energy properties. The sound mode $\w=\vLL|q|$ of the Luttinger liquid is replaced by a collective mode with dispersion $\w=v(q)|q|\sim |q|^{1+\sig/2}$ ($\w\sim |q|\sqrt{-\ln|q|}$ for $\sig=0$) and the compressibility $\kappa=\lim_{q\to 0} K(q)/\pi v(q)$ vanishes. Algebraic superfluid correlations are suppressed whereas translation invariance is spontaneously broken by the formation of a Wigner crystal with period $1/\rho_0$: $\mean{\hat\rho(q=2m\pi\rho_0)}=\rho_{2m}\mean{e^{2m i\hat\varphi(x)}}\neq 0$ ($m$ integer); for $\sig=0$, the order is only quasi-long-range~\cite{not1}. The Wigner crystal has a nonzero charge stiffness $D=\lim_{q\to 0}v(q)K(q)=\vLL\KLL$ independent of the long-range interactions.  

From now on, we restrict ourselves to genuine long-range interactions, i.e., $\sig\leq 0$. A weak disorder contributes to the Hamiltonian a term 
\beq 
\hat H_{\rm dis} = \int dx \llbrace -\frac{1}{\pi} \eta \dx \hat\varphi + \rho_2 [ \xi^* e^{2i\hat\varphi} + \hc] \rrbrace , 
\eeq
where we distinguish the so-called forward ($\eta$) and backward ($\xi$) scatterings; their Fourier components are near $0$ and $\pm 2\pi\rho_0$, respectively~\cite{Giamarchi87,Giamarchi88}. The forward scattering potential $\eta$ can be eliminated by a shift of $\hat\varphi$, i.e., $\hat\varphi(x)\to \hat\varphi(x)+\alpha(x)$ with a suitable choice of $\alpha(x)$, and is therefore discarded in the following (it does, however, play a role in some of the correlation functions discussed below). The average over disorder can be done using the replica method, i.e., by considering $n$ copies of the model. Assuming that $\xi(x)$ is Gaussian distributed with zero mean and variance $\overline{\xi^*(x)\xi(x')}=(\calD/\rho_2^2)\delta(x-x')$ (an overline indicates disorder averaging), we obtain the following low-energy Euclidean action (after integrating out the field $\theta$), 
\begin{multline}
S[\{\varphi_a\}] = \half \sum_{Q,a} \varphi_a(-Q) \left( Z_x q^2 f_q + \frac{\w^2}{\pi\vLL\KLL} \right) \varphi_a(Q) \\  
- \calD \sum_{a,b} \int dx \int_0^\beta d\tau\, d\tau' \cos[2\varphi_a(x,\tau)-2\varphi_b(x,\tau')] ,
\label{action}
\end{multline}
where $\varphi_a(x,\tau)$ is a bosonic field with $\tau\in[0,\beta]$ an imaginary time ($\beta=1/T\to\infty$), and $a,b=1\cdots n$ are replica indices. We use the notation $Q=(q,i\w)$ with $\w\equiv \wn=2n\pi T$ ($n$ integer) is a Matsubara frequency. In Eq.~(\ref{action}), $Z_xf_q=v(q)/\pi K(q)$ and in the following we use the low-momentum approximation $Z_xf_q\simeq \vLL/\pi\KLL+Z_x|q|^\sig$ (or $Z_xf_q\simeq \vLL/\pi\KLL+Z_x\ln|\Lambda/q|$ for $\sig=0$) valid when $|q|a\ll 1$. We can now identify two characteristic length scales. The first one, $L_x=(Z_x\pi\KLL/\vLL)^{1/\sig}$ is a crossover length beyond which the long-range potential $\calV_\sig$ dominates over the short-range interactions. The second one, the Larkin length $L_c\sim (Z_x^2/\calD)^{1/(3+2\sig)}$, signals the breakdown of perturbation theory with respect to disorder~\cite{not5}. The divergence of $L_c$ when $\sig\to-3/2$ suggests, as will be confirmed below, that the Wigner crystal is stable when $\sig<-3/2$.  

Most physical quantities can be obtained from the partition function $\calZ[\{J_a\}]$ or, equivalently, from the effective action (or Gibbs free energy) 
\begin{equation}
\Gamma[\{\phi_a\}] = - \ln\calZ[\{J_a\}] + \sum_a \int dx \inttau \,J_a \phi_a 
\end{equation}
defined as the Legendre transform of the free energy $-\ln\calZ[\{J_a\}]$. Here $J_a$ is an external source which couples linearly to the field $\varphi_a$ and allows us to obtain the expectation value $\phi_a(x,\tau)=\mean{\varphi_a(x,\tau)}=\delta\ln\calZ[\{J_f\}]/\delta J_a(x,\tau)$. We compute $\Gamma[\{\phi_a\}]$ using a Wilsonian nonperturbative FRG approach~\cite{Berges02,Delamotte12,Dupuis_review} where fluctuation modes are progressively integrated out. In practice we consider a scale-dependent effective action $\Gamma_k[\{\phi_a\}]$ which incorporates fluctuations with momenta (and frequencies) between a running momentum scale $k$ and the UV scale $\Lamb$. The effective action of the original model, $\Gamma_{k=0}[\{\phi_a\}]$, is obtained when all fluctuations have been integrated out whereas $\Gamma_\Lamb[\{\phi_a\}]=S[\{\phi_a\}]$. $\Gamma_k$ satisfies a flow equation which allows one to obtain $\Gamma_{k=0}$ from $\Gamma_\Lamb$ but which cannot be solved exactly~\cite{Wetterich93,Ellwanger94,Morris94}. 

Following previous FRG studies of one-dimensional disordered boson systems~\cite{Dupuis19,Dupuis20,Dupuis20a}, we consider the following truncation of the effective action, 
\begin{equation}
\Gamma_k[\{\phi_a\}] = \sum_a \Gamma_{1,k}[\phi_a] - \half \sum_{a,b} \Gamma_{2,k}[\phi_a,\phi_b] ,  
\end{equation}
with the ansatz 
\begin{equation}
\begin{split} 
&\Gamma_{1,k}[\phi_a] = \half \sum_Q \phi_a(-Q) [ Z_{x} q^2 f_q + \Delta_k(i\w) ] \phi_a(Q) , \\ 
&\Gamma_{2,k}[\phi_a,\phi_b] = \int dx \inttau\, d\tau'\, V_k(\phi_a(x,\tau)-\phi_b(x,\tau')) , 
\end{split}
\label{ansatz}
\end{equation}
and the initial conditions $\Delta_\Lamb(i\w)=\w^2/\pi v_{\rm LL}K_{\rm LL}$ and $V_\Lamb(u)=2\calD \cos(2u)$. The $\pi$-periodic function $V_k(u)$ can be interpreted as a renormalized second cumulant of the disorder. The form of the ansatz~(\ref{ansatz}) is strongly constrained by the so-called statistical tilt symmetry (STS)~\cite{Schulz88,Dupuis20}. In particular the term $Z_x q^2 f_q$ is not renormalized and no other space-derivative terms can be generated. The self-energy $\Delta_k(i\w)$ is {\it a priori} arbitrary but satisfies $\Delta_k(i\w=0)=0$. It is convenient to define $k$-dependent velocity and Luttinger parameter from $Z_x=v_k/\pi K_k f_k$ and $\Delta_k(i\w)=Z_x\w^2 f_k/v_k^2+\calO(\w^4)$. In the absence of disorder, $\Gamma_k[\{\phi_a\}]=S[\{\phi_a\}]$ and one has $v_k\sim f_k^{1/2}$ and $K_k\sim f_k^{-1/2}$ in agreement with the momentum-dependent quantities $v(q)$ and $K(q)$ [Eqs.~(\ref{Kqvq})]. 

$\Gamma_{1,k}$ and $\Gamma_{2,k}$ contain all the necessary information to characterize the ground state of the system. From the disorder-averaged density-density correlation function 
\beq 
\chi_{\rho\rho}(q,i\w) = \frac{q^2/\pi^2}{Z_x q^2 f_q + \Delta_{k=0}(i\w)} ,
\eeq 
we deduce that the compressibility 
\beq
\kappa = \lim_{q\to 0} \chi_{\rho\rho}(q,0) = \lim_{q\to 0} \frac{1}{\pi^2 Z_x f_q} = 0
\eeq 
vanishes so that the system remains incompressible in the presence of disorder. The determination of the conductivity $\sig(\w)=\lim_{q\to 0}(-i\w/q^2)\chi_{\rho\rho}(q,\w+i0^+)$ requires to determine the self-energy $\Delta_k(i\w)$ whose low-frequency behavior depends on $V_k(u)$. Incidentally the importance of disorder is best characterized by the dimensionless disorder correlator defined by 
$\delta_k(u) = - K_k^2 V_k''(u)/v_k^2 k^3$. We refer to the Supplemental Material for more details about the implementation of the FRG approach and the derivation of the flow equations for $\Delta_k(i\w)$ and $\delta_k(u)$~\cite{not2}.

\paragraph{FRG flow and phase diagram} 

\begin{figure}
\centerline{\includegraphics[width=8cm]{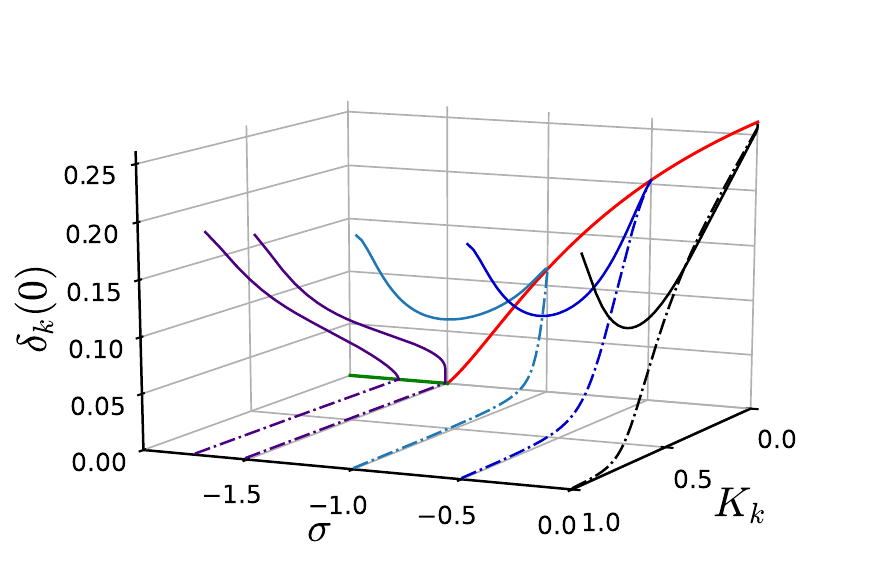}}
\caption{Flow trajectories $(K_k,\delta_k(0))$ for various values of $\sig$. The solid and dash-dotted lines are obtained for different values of the disorder strength. The red solid line for $-3/2<\sig\leq 0$ corresponds to the Mott-glass fixed point defined by $K^*=0$ and $\delta^*(u)$ [Eq.~(\ref{deltaFP})]. Disorder is irrelevant for $\sig<-3/2$ and the solid green line corresponds to the Wigner-crystal fixed point.} 
	\label{fig_flowdiag} 
\end{figure}

\begin{figure}
\centerline{\includegraphics[width=7.5cm]{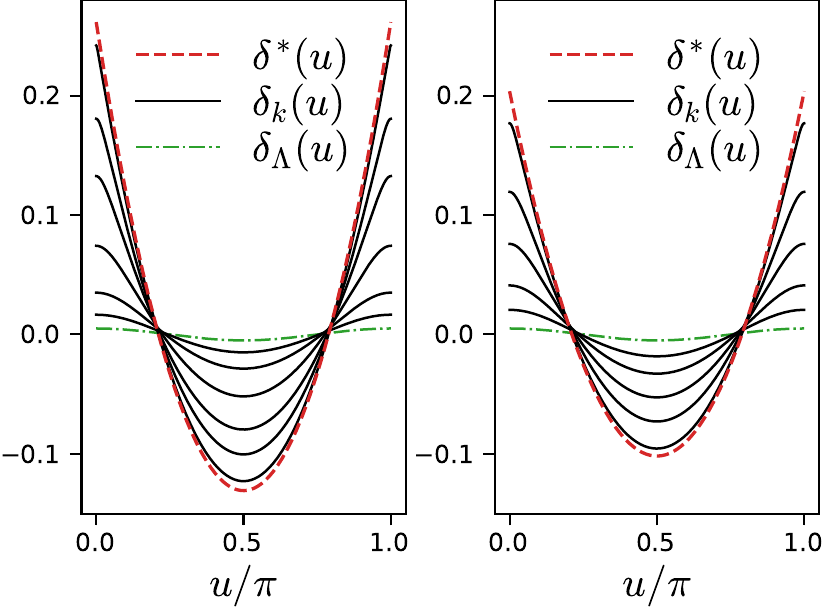}}
\caption{Disorder correlator $\delta_k(u)$ for various values of $k$ and $\sig=0$ (left), $\sig=-0.5$ (right). The green dash-dotted curve shows the initial condition $\delta_\Lambda(u)\propto\cos(2u)$ and the red dashed one the fixed-point solution~(\ref{deltaFP}).}
	\label{fig_delta} 
\end{figure}

By solving numerically the flow equations, we find that for $\sig>-3/2$ the flow trajectories are attracted by a fixed point characterized by a vanishing Luttinger parameter $K_k\sim k^{-\sig/2+\theta}\to 0$ (Fig.~\ref{fig_flowdiag}). The velocity behaves as $v_k\sim k^{\theta+\sig/2}$ and vanishes in the limit $k\to 0$ if $\sig>-2\theta$ but diverges (as in the Wigner crystal) if $\sig<-2\theta$. Whether the latter case actually occurs (which requires $\theta<3/4$ since $\sig>-3/2$ in the Mott glass) depends on the value of $\theta$ which, for reasons explained in Ref.~\cite{Dupuis20}, cannot be accurately determined from the flow equations. The charge stiffness $D_k=v_kK_k\sim k^{2\theta}$ vanishes for $k\to 0$ and the system is insulating. 

On the other hand, the disorder correlator $\delta_k(u)$ reaches a nontrivial fixed point in the limit $k\to 0$ when $\sig>-3/2$ (see Fig.~\ref{fig_delta}):  
\beq
\delta^*(u) = \frac{3+2\sig}{6\pi \bar l_2} \left[ \left( u-\frac{\pi}{2} \right)^2 - \frac{\pi^2}{12} \right] 
\quad (u\in [0,\pi]) ,
\label{deltaFP}
\eeq 
where $\bar l_2$ is a nonuniversal constant. Apart from the $\sig$-dependent prefactor, $\delta^*(u)$ is identical to the fixed-point solution in the Bose-glass phase~\cite{Dupuis19,Dupuis20}. It exhibits cusps at $u=n\pi$ ($n$ integer). For any nonzero momentum scale this cusp singularity is rounded into a quantum boundary layer (QBL) as shown in Fig.~\ref{fig_delta}: For $u$ near $n\pi$, $\delta_k(n\pi)-\delta_k(u)\propto |u-n\pi|$ except in a boundary layer of size $|u-n\pi|\sim K_k$, and the curvature $|\delta''_k(n\pi)|\sim 1/K_k\sim k^{-\theta+\sig/2}$ diverges when $k\to 0$. The cusp singularity and the QBL describes the physics of rare low-energy metastable states and their coupling to the ground state by quantum fluctuations~\cite{Dupuis19,Dupuis20}. This is characteristic of disordered systems with glassy properties~\cite{Balents96}. 

The behavior of the self-energy $\Delta_k(i\w)$ when $\sig>-3/2$ is also reminiscent of the Bose-glass phase. For small $k$, there is a frequency regime where $\Delta_k(i\w)$ is compatible with a linear dependence $A+B|\w|$, which implies that the real part of the conductivity  
\begin{align}
\sig(\w) &= -\frac{i\w}{\pi^2 \Delta_{k=0}(\w+i0^+)} \nonumber \\ 
&= \frac{1}{\pi A^2}(-iA\w+B\w^2) + \calO(\w^3) 
\end{align}
vanishes as $\w^2$~\cite{not6}. However, when $2-\theta+3\sig/2$ becomes negative, which necessarily occurs when $\sig$ varies between 0 and $-3/2$ since $\theta>0$, the constant $A$ grows and seems to diverge for $k\to 0$. This could indicate that the conductivity vanishes with an exponent larger than 2: $\Re[\sig(\w)]\ll \w^2$~\cite{not2}. Thus, for $\sig>-3/2$, we essentially recover the physical properties of the Bose-glass phase with the notable exception that the compressibility vanishes: The ground state is a Mott glass. 

In the Mott glass, the backward scattering destroys the long-range crystalline order: $\mean{\hat\rho(q=2\pi\rho_0)}= \rho_2 \mean{e^{ 2i\hat\varphi(x)}}=0$, and the corresponding correlation function $\chi(x)=\mean{e^{2i\hat\varphi(x)} e^{-2i\hat\varphi(0)}}$ decays algebraically. Taking into account the forward scattering, we find~\cite{not2} 
\beq 
\chi(x) \sim \llbrace \begin{array}{lll} 
	\frac{e^{-C|x|^{1+2\sig}}}{|x|^{\gamma_\sig}} & \mbox{if} & \sig> -1/2 , \\  
    \frac{1}{|x|^{\gamma_\sig}} & \mbox{if} & \sig< -1/2 
\end{array} \right. 
\eeq 
($C$ is a positive constant), where $\gamma_\sig = \pi^2 (3+2\sig)/9 +\theta-\sig/2$. Forward scattering is relevant for $\sig>-1/2$ and yields an exponential suppression of crystalline order but becomes irrelevant for $\sig<-1/2$~\cite{not2}. 

\begin{table}
	\caption{Some of the physical properties of the phases shown in the phase diagrams of Fig.~\ref{fig_phase_dia}: crystalline order, compressibility $\kappa$ and low-frequency optical conductivity $\sig(\w)$ (QLRO stands for quasi-long-range order).} 
	\begin{tabular}{lccc}
		\hline \hline
		& crystallization & $\kappa$ & $\Re[\sig(\w)]$  \\ 
		\hline
		Luttinger liquid $\sig>0$ & no & $>0$ & $D\delta(\w)$  \\
		Wigner crystal $\sig=0$ & QLRO & 0    & $D\delta(\w)$  \\ 
		Wigner crystal $-2<\sig<0$ & LRO & 0    & $D\delta(\w)$  \\ 
		Wigner crystal $\sig=-2$ & LRO & 0    & gapped  \\ 
		Bose glass        & no & $>0$ & $\w^2$  \\
		Mott glass        & no & 0    & $\w^2$  \\
		\hline
	\end{tabular}
	\label{table_phases}
\end{table}

When $\sig<-3/2$, both forward and backward scatterings are irrelevant and the Wigner crystal is stable against a weak disorder as shown by the flow trajectories in Fig.~\ref{fig_flowdiag}. Thus, for sufficiently long-range interactions, the Wigner crystal is sufficiently rigid to survive the detrimental effect of disorder. The case $\sig=-2$ (disordered Schwinger model) requires a separate study since $Z_xq^2 f_q$ does not vanish for $q\to 0$. Although there are contradicting results in the literature regarding the possible existence of a Mott glass in the disordered Schwinger model~\cite{Giamarchi01,Chou18,Dupuis20a}, our results regarding the stability of the Wigner crystal against disorder when $-2<\sig<-3/2$ are in line with a recent FRG study predicting the absence of a Mott glass when $\sig=-2$, the ground state being similar to a Mott insulator (vanishing compressibility and gapped conductivity)~\cite{Dupuis20a}. 

The phase diagram of a one-dimensional disordered Bose fluid with the long-range interaction potential $\calV_\sig$ [Eq.~(\ref{Vsig})] is shown in Fig.~\ref{fig_phase_dia}. In the absence of disorder, the ground state is a Luttinger liquid for effectively short-range interactions ($\sig>0$) and a Wigner crystal for genuine long-range interactions ($\sig\leq 0$). The Luttinger liquid is unstable against infinitesimal disorder and becomes a Bose glass when the Luttinger parameter satisfies $\KLL<3/2$ (with $\KLL=\lim_{q\to 0}K(q)$ including the effect of the potential $\calV_\sig$)~\cite{Giamarchi87,Giamarchi88}. On the other hand disorder transforms the Wigner crystal into a Mott glass when $\sig>-3/2$. Some of the physical properties of these various phases are summarized in Table~\ref{table_phases}.

\paragraph{Conclusion.}
We have shown that a one-dimensional disordered Bose fluid with long-range interactions exhibits a rich phase diagram which includes the long-sought Mott-glass phase. Since the Hamiltonian studied in this Letter also describes the charge degrees of freedom of fermions, a similar phase diagram is expected for a one-dimensional Fermi fluid. 

On the experimental side, long-range interactions have been realized in various cold-atom systems, e.g. trapped ions~\cite{Britton12,Islam13,Richerme14} or dipolar quantum gases~\cite{Baranov12}, and we may hope that one-dimensional quantum fluids with long-range interactions will be realized in the near future. Of particular interest are cold-atom systems in an optical lattice and using an optical cavity to realize the Hubbard model with an additional infinite-range (cavity-mediated) interaction~\cite{Landig16,Botzung19a}. In the presence of disorder this system, in one dimension, would be described by the low-energy model studied in this Letter. But the scaling of the long-range interaction with the system size, the so-called Kac prescription~\cite{Kac63}, prevents a direct comparison with the results of this Letter~\cite{Botzung19a}. On the other hand we note that the Schwinger model has already been realized~\cite{Martinez16} and allows for a check of our prediction regarding the stability of the Wigner crystal when $\sig=-2$.

\paragraph{Acknowledgment.} We thank N. Defenu, J. Beugnon and P. Viot for useful comments on the experimental realization of long-range interactions in cold-atom systems. 

\input{MG_final.bbl}


\clearpage

\onecolumngrid

\begin{center}
	\textbf{\large  One-dimensional disordered Bose fluid with long-range interactions \\ 
		[.3cm] -- Supplemental Material --} \\ 
	[.4cm] Romain Daviet and Nicolas Dupuis \\[.1cm]
	{\itshape Sorbonne Universit\'e, CNRS, Laboratoire de Physique Th\'eorique de la Mati\`ere Condens\'ee, LPTMC, F-75005 Paris, France \\}
	(Dated: December 12, 2023)\\[0.5cm]
\end{center}

\setcounter{equation}{0}
\setcounter{figure}{0}
\setcounter{table}{0}
\setcounter{page}{1}
\renewcommand{\theequation}{S\arabic{equation}}
\renewcommand{\thefigure}{S\arabic{figure}}	
\renewcommand{\bibnumfmt}[1]{[S#1]}
\renewcommand{\citenumfont}[1]{S#1}

In the Supplemental Material, we discuss in detail the one-dimensional Bose fluid with long-range interactions with and without disorder, and present the nonperturbative functional renormalization-group (FRG) approach used to determine the phase diagram.

\section{I.\;\; Pure Bose fluid} 

\subsection{A.\;\; Long-range interaction potential} 

We consider a long-range interaction potential $\calV_\sig(x)$ defined by 
\beq
\calV_\sig(x) = \llbrace 
\begin{array}{lcl} 
	\dfrac{e^2}{(x^2 + a^2)^{(1+\sig)/2}} & \mbox{if} & -1<\sig , \\ [3ex] 
	-e^2 \ln|x/a| & \mbox{if} & \sig=-1 , \\ [1.5ex] 
	-e^2 |x|^{-\sig-1} & \mbox{if} & -2\leq \sig <-1 , 
\end{array}
\right.
\label{eqsm13}
\eeq
where the particle ``charge'' $e$ is introduced to make these definitions dimensionally correct. We assume the presence of a  uniform background of charge $-e$ to make the system globally neutral so that the long-range part of the Hamiltonian reads $\half \int dxdx' (\hat\rho(x)-\rho_0) \calV_\sig(x-x')(\hat\rho(x')-\rho_0)$, which leads to the bosonized Hamiltonian~(3) in the main text. The short-distance cutoff $a$ is necessary to make the Fourier transform $\calV_\sig(q)$ of the potential $\calV_\sig(x)$ well defined when $\sig\geq 0$. $\calV_0(x)$ is the Coulomb potential while $\calV_{-2}$ corresponds to the Schwinger model where the particles interact {\it via} a $(1+1)$-dimensional gauge field~\cite{Schwinger62sm,Coleman76sm}. $\calV_\sig(q)$ is given by 
\beq 
\calV_\sig(q) = \llbrace 
\begin{array}{lll} 
	\frac{2^{1-\frac{\sig}{2}}\sqrt{\pi}e^2}{\Gamma(\frac{1+\sig}{2})} \left|\dfrac{q}{a}\right|^\frac{\sig}{2} K_{-\frac{\sig}{2}}(a|q|) & \mbox{if} & -1<\sig, \\ [1.5ex] 
	\dfrac{\pi e^2}{|q|} +2\pi e^2 \ln(e^\gamma a) \delta(q) & \mbox{if} & \sig=-1 , \\ [3ex] 
	- 2 e^2 |q|^\sig \cos\left(\frac{\pi\sig}{2}\right) \Gamma(-\sig) & \mbox{if} & -2\leq \sig <-1 , 
\end{array}
\right.
\eeq
where $\gamma$ is the Euler constant and $K_{-\sig/2}$ the modified Bessel function of the second kind. We can eliminate the Dirac function in $\calV_{-1}(q)$ by choosing $a=e^{-\gamma}$. When $q\to 0$, the Fourier transformed potential $\calV_\sig(q)$ has a finite limit for $\sig>0$, whereas $\calV_0(q)\sim -e^2\ln|qa|$ in the Coulomb case and $\calV_\sig(q)\sim |q|^\sig$ for $\sig<0$.

\subsection{B.\;\; Bosonization and correlation functions}

Let us consider one-dimensional bosons interacting {\it via} a short-range potential and the long-range potential $\calV_\sig(x)$. In the bosonization formalism~\cite{Giamarchi_booksm}, one introduces two phase operators, $\hat\theta$ and $\hat\varphi$, which satisfy the commutation relations $[\hat\theta(x),\dy\hat\varphi(y)]=i\pi\delta(x-y)$ and are related to the boson operator by $\hat\psi(x)=e^{i\hat\theta(x)} \hat\rho(x)^{1/2}$ and 
\beq 
\hat\rho(x) = \rho_0 - \frac{1}{\pi} \dx\hat\varphi(x) + 2 \sum_{m=1}^\infty \rho_{2m} \cos(2m \pi \rho_0 x-2m\hat\varphi(x)) , 
\eeq 
where $\rho_0$ is the mean density and the $\rho_{2m}$'s are nonuniversal parameters that depend on microscopic details. At low energies, the Hamiltonian can then be written as 
\begin{equation}
\hat H = \sum_q \frac{v(q) q^2}{2\pi} \llbrace \frac{1}{K(q)} \hat\varphi(-q)\hat\varphi(q) + K(q) \hat\theta(-q) \hat\theta(q) 
\rrbrace,
\label{eqsm0}
\end{equation}
where a UV momentum cutoff $\Lambda$ is implied. The momentum-dependent velocity and Luttinger parameter are defined by 
\beq 
v(q)K(q) = \vLL\KLL , \qquad 
\frac{v(q)}{K(q)} = \frac{\vLL}{\KLL} + \frac{\calV_\sig(q)}{\pi} ,
\eeq  
and $\vLL$ and $\KLL$ are the parameters associated with the short-range interactions. When $\sig>0$, since $\calV_\sig(q)$ has a finite limit for $q\to 0$, $v(q=0)$ and $K(q=0)$ are finite. In that case the potential $\calV_\sig$ is effectively short-range and leads to a mere renormalization of $\vLL$ and $\KLL$ (in the following, we assume that the effect of the potential $\calV_\sig$, when $\sig>0$, is taken into account by redefining $\vLL$ and $\KLL$); the ground state remains a Luttinger liquid. By contrast, when $\sig\leq 0$, in the small-momentum limit $\calV_\sig(q)\sim |q|^\sig$ so that $v(q)\sim |q|^{\sig/2}$ and $K(q)\sim |q|^{-\sig/2}$ are determined by the long-range part of the interactions (for $\sig=0$, $|q|^\sig$ should be interpreted as $-\ln|q|$). The spectrum is given by $\w_q=v(q)|q|$; the sound mode with linear dispersion $\w_q=\vLL|q|$ that exists in the Luttinger liquid is therefore replaced by a collective mode with dispersion $\w_q\sim |q|^{1+\sig/2}$ (or $\w_q\sim |q|\sqrt{-\ln|q|}$ for Coulomb interactions) in the presence of long-range interactions ($-2<\sig\leq 0$). The spectrum is gapped for $\sig=-2$ (Schwinger model): $\w_q^2=\vLL^2q^2+2e^2\vLL\KLL/\pi$. 

From the Hamiltonian~(\ref{eqsm0}) one easily obtains the propagators of the fields $\hat\varphi$ and $\hat\theta$, 
\beq
G_{\varphi\varphi}(q,i\w) = \frac{\pi v(q) K(q)}{\w^2+v(q)^2 q^2} , \qquad 
G_{\theta\theta}(q,i\w) = \frac{\pi v(q)/K(q)}{\w^2+v(q)^2 q^2} ,
\label{eqsm6}
\eeq 
where $\w\equiv \wn=2n\pi T$ ($n$ integer) is a Matsubara frequency (we drop the index $n$ since we consider only the $T=0$ limit where $\wn$ becomes a continuous variable). 

Consider now the order parameters $\Delta_m=\mean{e^{2im\hat\varphi(x)}}$ ($m$ integer) associated with a spontaneous modulation of the density with period $1/\rho_0$:  
\beq 
\mean{\hat\rho(x)} = \rho_0 + 2 \sum_{m=1}^\infty \rho_m \Delta_m \cos(2m\pi\rho_0x) .
\eeq
The Hamiltonian being quadratic, one easily obtains 
\beq
\Delta_m = e^{-2m^2\mean{\hat\varphi(x)^2}} \quad \mbox{with} \quad  
\mean{\hat\varphi(x)^2} = \intinf \frac{dq}{2\pi} \intw G_{\varphi\varphi}(q,i\w) .
\eeq 
For a Luttinger liquid ($\sig>0$), i.e., $v(q)\equiv\vLL$ and $K(q)\equiv\KLL$ for $q\to 0$, one finds that $\mean{\hat\varphi(x)^2}$ diverges and $\Delta_m=0$: The ground-state density is uniform. By contrast, when $\sig<0$, the small-momentum behavior $v(q)\sim|q|^{\sig/2}$ and $K(q)\sim |q|^{-\sig/2}$ makes $\mean{\hat\varphi^2}$ finite so that $\Delta_m\neq 0$: The ground state is a Wigner crystal. There is quasi-long-range order when $\sig=0$~\cite{Schulz93sm}. 

One can further characterize the ground state by computing the compressibility 
\beq 
\kappa = \lim_{q\to 0} \chi_{\rho\rho}(q,0) = \lim_{q\to 0} \frac{K(q)}{\pi v(q)} =  \lim_{q\to 0} \left( \frac{\pi\vLL}{\KLL} + \calV_\sig(q) \right)^{-1}   
\eeq 
($\chi_{\rho\rho}(q,i\w)=(q/\pi)^2 G_{\varphi\varphi}(q,i\w)$ is the long-wavelength density-density correlation function) and the charge stiffness (or Drude weight)
\beq 
D = \lim_{q\to 0} v(q) K(q) = \vLL \KLL 
\eeq  
defined as the weight of the Dirac peak in the optical conductivity 
\beq
\sig(\w) = \lim_{q\to 0} \frac{-i\w}{q^2} \chi_{\rho\rho}(q,\w+i0^+) = D \left[ \delta(\w) + \frac{i}{\pi} \calP \frac{1}{\w} \right] , 
\eeq
where $\calP$ denotes the principal part. The compressibility $\kappa=\KLL/\pi\vLL$ is finite in the Luttinger liquid ($\sig>0$) but vanishes in the Wigner crystal ($\sig\leq 0$). On the other hand, the charge stiffness $D=\vLL\KLL$ is always finite and fully determined by the short-range interactions. 

Let us finally consider the long-distance behavior of the superfluid correlation function $G(x)$, and the $q\simeq 2\pi\rho_0$ (connected) density correlation function $\chi(x)$, 
\beq
\begin{split}
	G(x) &=\mean{\hat\psi(x)\hat\psi^\dagger(0)}\simeq \rho_0 \mean{e^{i\hat\theta(x)} e^{-i\hat\theta(0)}} = \rho_0 e^{G_{\theta\theta}(x)-G_{\theta\theta}(0)} , \\
	\chi(x) &=\mean{e^{2i\hat\varphi(x)} e^{-2i\hat\varphi(0)}} - \mean{e^{-2i\hat\varphi(0)}}^2 = e^{-4G_{\varphi\varphi}(0)]} [ e^{4 G_{\varphi\varphi}(x)} - 1 ], 
\end{split}
\label{eqsm12}
\eeq
where the equal-time correlation functions $G_{\varphi\varphi}(x)=\mean{\hat\varphi(x)\hat\varphi(0)}$ and $G_{\theta\theta}(x)=\mean{\hat\theta(x)\hat\theta(0)}$ can be obtained from Eqs.~(\ref{eqsm6}). The results are summarized in Table~\ref{table2_phases}. In the Luttinger liquid, the superfluid correlations dominate when $K>1/2$ whereas the $q\simeq 2\pi\rho_0$ density correlations are the leading ones when $K<1/2$. In the presence of Coulomb interactions ($\sig=0$), $\chi(x)$ ($G(x)$) decays slower (faster) than any power law; although there is no genuine long-range crystalline order, the ground state can be seen as a Wigner crystal~\cite{Schulz93sm}. For longer range interactions ($-2\leq\sig<0$), $G(x)$ decays as a (stretched) exponential while there is long-range crystalline order. 

\begin{table}
	\caption{Various ground states of a one-dimensional Bose fluid with (some of) their physical properties: Wigner crystallization (associated with the order parameter $\mean{e^{2i\hat\varphi(x)}}$), compressibility $\kappa$, low-frequency optical conductivity $\sig(\w)$ ($D$ denotes the charge stiffness or Drude weight), $q=2\pi\rho_0$ density-density correlation function $\chi(x)$ and superfluid correlation function $G(x)$ [Eqs.~(\ref{eqsm12})]. In the pure system, the ground state is either a Luttinger liquid (LL), if the potential $\calV_\sig$ [Eq.~(\ref{eqsm13})] is effectively short range ($\sig>0$) or a Wigner crystal (WC) in the presence of genuine long-range interactions ($\sig\leq 0$) (there is only quasi-long-range order (QLRO) for Coulomb interactions~\cite{Schulz93sm}). In the presence of disorder, the LL is unstable against the formation of a Bose glass (nonzero compressibility and gapless conductivity) when $\KLL<3/2$~\cite{Giamarchi87sm,Giamarchi88sm,Ristivojevic12sm,Ristivojevic14sm,Dupuis19sm,Dupuis20sm}, whereas the WC becomes a Mott glass (vanishing compressibility and gapless conductivity) if $\sig>-3/2$. The WC is stable for $\sig<-3/2$. For $\sig=-2$ (Schwinger model), the ground state exhibits long-range crystalline order, a vanishing compressibility and a gapped conductivity, and is very similar to a Mott insulator~\cite{Dupuis20asm}. [$\gamma_\sig=\pi^2(3+2\sig)/9+\theta-\sig/2$ and $C$ denotes a positive constant.] } 
	\begin{tabular}{lcccccc}
		\hline \hline
		& $\begin{array}{cc} \mbox{crystalline} \\ \mbox{order} \end{array}$ & $\kappa$ & $\Re[\sig(\w)]$ & $\chi(x)$ & $G(x)$ & $\begin{array}{c} \mbox{stability against} \\ \mbox{(infinitesimal) disorder} \end{array}$ \\ 
		\hline\hline
		\;LL ($\sig>0$)  & no & $>0$ & $D\delta(\w)$ & $|x|^{-2\KLL}$ & $|x|^{-1/2\KLL}$ & $\begin{array}{l} \mbox{stable if } \KLL>3/2 \\ \mbox{BG if } \KLL<3/2 \end{array}$ \\ \hline
		$\begin{array}{l} \mbox{WC  } (\sig=0) \\ \mbox{Coulomb interactions} \end{array}$   & QLRO & 0    & $D\delta(\w)$ & $e^{-C(\ln|x|)^{1/2}}$ & $e^{-C(\ln|x|)^{3/2}}$ & unstable (MG) \\ \hline
		\;WC ($-2<\sig<0$)   & LRO & 0    & $D\delta(\w)$ & $|x|^{\sig/2}$ & $e^{-C|x|^{-\sig/2}}$ & 
		$\begin{array}{l} \mbox{stable if } \sig<-3/2 \\ \mbox{MG if } \sig>-3/2 \end{array}$ \\ \hline 
		$\begin{array}{l} \mbox{WC  } (\sig=-2) \\ \mbox{Schwinger model} \end{array}$   & LRO & 0 & gapped & $e^{-C|x|} |x|^{-1/2}$ & $e^{-C|x|}$  & stable \\ \hline 
		$\begin{array}{l} \mbox{BG  }  \\ \mbox{short-range interactions} \end{array}$ & no & $>0$ & $\w^2$ & $e^{-C|x|}|x|^{-\gamma_0}$ & ? & $-$ \\ \hline
		\;MG  ($-3/2<\sig\leq 0$)  & no & 0    & $\w^2$ & $\begin{array}{ll} e^{-C|x|^{1+2\sig}}|x|^{-\gamma_\sig} & (\sig>-1/2) \\ |x|^{-\gamma_\sig} & (\sig<-1/2)  \end{array} $ & ? & $-$ \\
		\hline
	\end{tabular}
	\label{table2_phases}
\end{table}

\section{II.\;\; Disordered Bose fluid} 

From now on, we restrict ourselves to genuine long-range interactions, i.e., $\sig\leq 0$. In the presence of disorder, it is convenient to use the functional integral formalism. After integrating out the field $\theta$, one obtains the action 
\beq
S[\varphi;\xi,\eta] = \sum_{q,\w} \frac{v(q)}{2\pi K(q)} \left( q^2 + \frac{\w^2}{v(q)^2} \right) \varphi(-q,-i\w) \varphi(q,i\w) 
+ \int_{x,\tau} \llbrace - \frac{1}{\pi} \eta \dx \varphi + \rho_2 [ \xi^* e^{2i\varphi} + \cc ] \rrbrace ,
\label{eqsm8}
\eeq 
where $\varphi(x,\tau)$ is a bosonic field with $\tau\in [0,\beta]$ an imaginary time ($\beta=1/T\to\infty$). 
We use the notation $\int_{x,\tau}=\int dx \inttau$. The forward and backward scattering random potentials, $\eta$ and $\xi$, have Fourier components near 0 and $\pm 2\pi\rho_0$, respectively. The partition function 
\beq 
\calZ[J;\xi,\eta] = \int \calD[\varphi] \, e^{-S[\varphi;\xi,\eta] + \int_{x,\tau} J\varphi} 
\eeq 
is a functional of both the external source $J$ and the random potentials $\eta$ and $\xi$. The physics of the system is determined by the cumulants of the random functional $W[J;\xi,\eta]=\ln \calZ[J;\xi,\eta]$: 
\beq
\begin{split} 
	W_1[J_a] ={}& \overline{W[J_a;\xi,\eta]} , \\ 
	W_2[J_a,J_b] ={}& \overline{W[J_a;\xi,\eta] W[J_b;\xi,\eta]}  - \overline{W[J_a;\xi,\eta]} \; \overline{W[J_b;\xi,\eta]} , 
\end{split} 
\eeq   
etc., where an overline indicates disorder averaging. The first cumulant gives the disorder-averaged free energy $W_1[J_a=0]$ while the second one, $W_2$, can be seen as a renormalized disorder correlator and assumes a cuspy functional form when the physics is determined by low-energy metastable states~\cite{Balents96sm,Tarjus19sm}.

\subsection{A.\;\; Replica formalism}
\label{subsec_replica} 

The cumulants can be computed by considering $n$ copies (or replicas) of the system, each with a different external source, and performing the disorder averaging. Assuming that $\eta$ and $\xi$ are Gaussian distributed with zero mean and variances $\overline{\xi^*(x)\xi(x')}=(\calD/\rho_2^2)\delta(x-x')$, $\overline{\eta(x)\eta(x')}=2\pi^2\calF\delta(x-x')$, this leads to the partition function 
\beq
\calZ[\{J_a\}] = \overline{\prod_{a=1}^n \calZ[J_a;\xi,\eta]}  = \int\calD[\{\varphi_a\}] \, e^{-S[\{\varphi_a\}]} ,
\eeq 
with the replicated action 
\begin{align}
S[\{\varphi_a\}] ={}&  \half \sum_{q,\w,a} \left( Z_x q^2 f_q + \frac{\w^2}{\pi \KLL \vLL} \right) \varphi_a(-q,-i\w) \varphi_a(q,i\w) 
\nonumber \\ & - \sum_{a,b} \int_{x,\tau,\tau'} \Bigl\{ \calF \bigl(\dx\varphi_a(x,\tau)\bigr)\bigl(\dx\varphi_b(x,\tau')\bigr) + \calD \cos[2\varphi_a(x,\tau)-2\varphi_b(x,\tau')] 
\Bigr\} ,
\label{eqsm7}
\end{align}
where $a,b=1\cdots n$ are replica indices and $Z_xf_q=v(q)/\pi K(q)$. In the following we use the low-momentum approximation $Z_xf_q\simeq\vLL/\pi\KLL+Z_x|q|^\sig$ (or $Z_xf_q\simeq \vLL/\pi\KLL+Z_x\ln|\Lambda/q|$ if $\sig=0$). The functional
\beq
W[\{J_a\}] = \ln \calZ[\{J_a\}] = \sum_a W_1[J_a] + \half \sum_{a,b} W_2[J_a,J_b] + \third \sum_{a,b,c} W_3[J_a,J_b,J_c] + \cdots 
\eeq 
is simply related to the cumulants $W_i$ of the random functional $W[J;\xi,\eta]$~\cite{Dupuis20sm}.

Before describing the computation of the cumulants $W_1$ and $W_2$ in the FRG formalism, let us discuss the stability of the Wigner crystal against disorder. Since the phase field $\varphi$ must have a vanishing scaling dimension, the non-disordered part of the action, $S_0$, has scaling dimension $\sig/2$ and the dynamical critical exponent in the Wigner crystal phase is $z=1+\sig/2$. On the other hand the scaling dimensions of the forward and backward scattering parts of the action are obtained from the scaling dimension $[S_{\rm dis}]=[S_0^2]=2[S_0]$ of the disordered part of the action, which gives $[\calF]=-1+2z+\sig=1+2\sig$ and $[\calD]=1+2z+\sig=3+2\sig$, respectively~\cite{notsm2}. We deduce that forward scattering is relevant only if $\sig>-1/2$ while backscattering is relevant if $\sig>-3/2$: The Wigner crystal is stable against disorder when $\sig<-3/2$.

Finally we note that the forward scattering can be eliminated from the action $S[\varphi;\xi,\eta]$ [Eq.~(\ref{eqsm8})] by an appropriate shift of the field, 
\beq 
\varphi(x,\tau) \to \varphi(x,\tau) + \alpha(x) \quad \mbox{with} \quad \alpha(q) = - i \frac{K(q)\eta(q)}{v(q)q} 
\simeq - \frac{i}{\pi Z_x} \frac{\eta(q) \sgn(q)}{|q|^{1+\sig}} ,
\label{eqsm9}
\eeq 
where the last expression corresponds to the small-momentum limit. The shift~(\ref{eqsm9}) does not leave the backward scattering term invariant, but this can be compensated by a redefinition of the random potential $\xi$, namely $\xi(x)\to\xi(x) e^{2i\alpha(x)}$, without changing its variance. This allows us to set $\calF=0$ in the replicated action~(\ref{eqsm7}). The compressibility and the conductivity are not modified by the shift, i.e., they keep the same expression in terms of the propagator $G_{\varphi\varphi}(q,i\w)$ of the (shifted) field $\varphi$. But the $q=2\pi\rho_0$ density correlation function $\chi(x)$ is multiplied by a factor $e^{2i[\alpha(x)-\alpha(0)]}$. After disorder averaging, we thus obtain
\beq 
\chi(x) = \Bigl[ \overline{\mean{e^{2i\varphi(x,0)} e^{-2i\varphi(0,0)}}} - \overline{\mean{e^{2i\varphi(x,0)}} \mean{e^{-2i\varphi(0,0)}}} \Bigr] e^{4\overline{\alpha(x)\alpha(0)}-4\overline{\alpha(0)^2}},
\eeq 
where $\overline{\alpha(x)\alpha(0)} - \overline{\alpha(0)^2}$ is given by 
\beq 
\frac{2\calF}{Z_x^2} \int \frac{dq}{2\pi} \frac{\cos(qx)-1}{|q|^{2+2\sig}}  
=  \frac{2\calF}{\pi Z_x^2} \llbrace 
\begin{array}{lll}
	- \Gamma(-1-2\sig)\sin(\pi\sig) |x|^{1+2\sig} & \mbox{if} & -\half <\sig<0 \\
	-  \ln|x| & \mbox{if} & \sig=-\half  
\end{array}
\right. 
\eeq 
(for $|x|\gg 1/\Lambda$) and goes to a finite limit for $|x|\to\infty$ when $\sig<-1/2$ in agreement with the irrelevance of the forward scattering in that case.

\subsection{B.\;\; FRG flow equations}

To implement the nonperturbative FRG approach~\cite{Berges02sm,Delamotte12sm,Dupuis_reviewsm} we add to the action the infrared regulator term 
\begin{equation}
\Delta S_k[\{\varphi_a\}] = \half \sum_{q,\w,a} \varphi_a(-q,-i\w) R_k(q,i\w) \varphi_a(q,i\w) 
\label{DeltaS}
\end{equation}
such that fluctuations are smoothly taken into account as $k$ is lowered from the microscopic scale $\Lamb$ down to 0. The partition function of the replicated system, 
\begin{equation}
\calZ_k[\{J_a\}] = \int \calD[\{\varphi_a]\}\, e^{-S[\{\varphi_a\}]-\Delta S_k[\{\varphi_a\}] + \sum_a \inttau \int dx  J_a \varphi_a } , 
\end{equation}
is now $k$ dependent. The main quantity of interest in the FRG formalism is the scale-dependent effective action 
\begin{equation}
\Gamma_k[\{\phi_a\}] = -\ln\calZ_k[\{J_a\}] + \sum_a \inttau \int dx \, J_a \phi_a - \Delta S_k[\{\phi_a\}] ,
\end{equation}
defined as a modified Legendre transform of $\ln\calZ_k[\{J_a\}]$ which includes the subtraction of $\Delta S_k[\{\phi_a\}]$. Here $\phi_a(x,\tau)=\mean{\varphi_a(x,\tau)}=\delta\ln\calZ_k[\{J_f\}]/\delta J_a(x,\tau)$ is the expectation value of the phase field. Assuming that all fluctuations are frozen by $\Delta S_\Lamb$, $\Gamma_\Lamb[\{\phi_a\}]=S[\{\phi_a\}]$. On the other hand the effective action of the original model is given by $\Gamma_{k=0}$ since $\Delta S_{k=0}=0$. The FRG approach aims at determining $\Gamma_{k=0}$ from $\Gamma_\Lamb$ using Wetterich's equation~\cite{Wetterich93sm,Ellwanger94sm,Morris94sm}
\begin{equation}
\dt \Gamma_k[\{\phi_a\}] = \half \Tr \Bigl\{ \dt R_k \bigl( \Gamma_k^{(2)}[\{\phi_a\}] + R_k )^{-1} \Bigr\} , 
\label{eqsm10}
\end{equation}
where $t=\ln(k/\Lamb)$ is a (negative) RG ``time'' and the trace involves a sum over momenta and frequencies as well as the replica index. Equation~(\ref{eqsm10}) cannot be solved exactly and we rely on the following truncation, 
\beq 
\Gamma_k[\{\phi_a\}] = \sum_a \Gamma_{1,k}[\phi_a] - \half \sum_{a,b}\Gamma_{2,k}[\phi_a,\phi_b] ,
\label{truncation}  
\eeq 
where the ansatz for $\Gamma_{1,k}$ and $\Gamma_{2,k}$ is given in the main text. $\Gamma_{1,k}[\phi_a]$ is the Legendre transform of the first cumulant $W_1[J_a]$ and contains all information about the thermodynamics, whereas $\Gamma_{2,k}[\phi_a,\phi_b]$ can be directly identified with the second cumulant $W_2[J_a,J_b]$~\cite{Dupuis20sm}. The truncation~(\ref{truncation}) has appeared in various models and there are no known examples where it has been shown to fail. In particular, it has been used in the nonperturbative FRG approach to the $d$-dimensional random-field Ising/O($N$) model where it gives a consistent and unified description of the equilibrium behavior in the whole $(N,d)$ diagram, and yields an estimate of the critical exponents in very good agreement with computer simulations in all dimensions~\cite{Tarjus19sm}. Furthermore, for the random elastic manifold model, which corresponds to the classical limit (in $d$ dimensions) of the boson model discussed in the manuscript, inclusion of the third cumulant $\Gamma_3[\phi_a,\phi_b,\phi_c]$ does not lead to any qualitative change and the truncation~(\ref{truncation}) appears semi-quantitatively correct~\cite{Balog19asm}. This strongly supports the validity of the approximation~(\ref{truncation}) even though a more stringent test would be to include $\Gamma_3$ in the one-dimensional quantum model. On a more qualitative level, we believe that the success of the truncation~(\ref{truncation}) comes from the fact that the second-order cumulant $\Gamma_2$ reflects the existence of the metastable states that determine the low-energy physics~\cite{Dupuis20sm}.

In practice, we choose the regulator function in the form
\begin{equation} 
R_k(q,i\w) = ( Z_x q^2 f_q + \Delta_k(i\w)) r\left( \frac{Z_x q^2 f_q + \Delta_k(i\w)}{Z_xk^2 f_k} \right) .
\label{eqsm11} 
\end{equation}
In Eq.~(\ref{eqsm11}), $r(x)=\alpha/(e^x-1)$ with $\alpha$ a parameter of order unity. Thus $\Delta S_k[\{\varphi_a\}]$ suppresses fluctuations such that $q^2\ll k^2$ and $\Delta_k(i\w)\ll Z_xk^2f_k$ but leaves unaffected those with $q^2\gg k^2$ or $\Delta_k(i\w)\gg Z_xk^2f_k$. The $k$-dependent Luttinger parameter and velocity, $K_k$ and $v_k$, are defined from $Z_x=v_k/\pi K_kf_k$ and $\Delta_k(i\w)=Z_x\w^2f_k/v^2_k+\calO(\w^4)$ (the presence of the regulator term $\Delta S_k$ in the action ensures that $\Delta_k(i\w)$ is an analytic function near $\w=0$). 

The derivation of the flow equations in a one-dimensional disordered Bose fluid with short-range interactions can be found in Ref.~\onlinecite{Dupuis20sm}. In the case of a long-range interaction potential, a similar derivation leads to
\begin{equation}
\begin{split} 
\dt\delta_k(u) ={}& -(3+2\sig_k) \delta_k(u) - K_k l_1 \delta''_k(u) 
+ \pi \bar l_2 [ \delta_k''(u) (\delta_k(u)-\delta_k(0)) + \delta'_k(u)^2 ] , \\ 
\dt \tilde\Delta_k(i\tw) ={}& - (2+\sig_k) \tilde\Delta_k(i\tw) + z_k \tw \partial_{\tw} \tilde\Delta_k(i\tw)
- \pi \delta''_k(0) [ \bar l_1(i\tw) - \bar l_1(0) ] , \\ 
\dt K_k ={}& \left( \theta_k - \frac{\sig_k}{2} \right) K_k , \qquad
\dt v_k = (z_k-1) v_k , 
\end{split}
\label{rgeq} 
\end{equation}
where   
\begin{equation}
z_k = 1 + \frac{\sig_k}{2} + \theta_k , \quad \sig_k= \dt \ln f_k, \quad \theta_k = \frac{\pi}{2} \delta''_k(0) \bar m_\tau ,
\label{thetak} 
\end{equation}
and $\delta_k(u)=-K_k^2 V_k''(u)/v_k^2 k^3$, $\tDelta_k(i\tw)=\Delta_k(i\w)/Z_xk^2f_k$. Here and below, $\tilde q=q/k$ and $\tw=\w/v_kk$ are dimensionless momentum and frequency variables. $\sig_k\to\sig$ when $\sig<0$ whereas, when $\sig=0$, $\sig_k\sim 1/\ln k\to 0$ for $k\to 0$.

The threshold functions appearing in Eqs.~(\ref{rgeq}) and (\ref{thetak}) are defined by
\begin{equation}
l_n = n \int_0^\infty d\tilde q \int_{-\infty}^\infty \frac{d\tw}{2\pi} \dt \tilde R_k(\tilde q,i\tw) 
\tilde G_k(\tilde q,i\tw)^{n+1} , \quad 
\bar l_n(i\tw) = n \int_0^\infty d\tilde q \,\dt \tilde R_k(\tilde q,i\tw) 
\tilde G_k(\tilde q,i\tw)^{n+1} ,
\label{threshold}
\end{equation}
$\bar l_n\equiv \bar l_n(0)$, and $\bar m_\tau = \partial_{\tw^2} \bar l_1(\tw) \bigl|_{\tw=0}$, where 
\begin{align} 
\tilde G_k(\tilde q,i\tw) ={}&  \frac{1}{(\tilde q^{2}f_q/f_k + \tilde\Delta_k)(1+r)} , \nonumber \\  
\dt \tilde R_k(\tilde q,i\tw)  ={}& (2+\sig_k) \tilde\Delta_k r 
- (2+\sig_k) \Bigl(\tilde q^{2} \frac{f_q}{f_k} + \tilde\Delta_k \Bigr) \tilde q^{2}\frac{f_q}{f_k} r' + (\dt\tilde\Delta_k - z_k \tw \partial_\tw \tilde\Delta_k)\Bigl[r + \Bigl(\tilde q^2 \frac{f_q}{f_k} + \tilde\Delta_k \Bigr) r' \Bigr] , 
\end{align}
with $r\equiv r(\tilde q^{2}f_q/f_k+\tilde\Delta_k)$ and $\tilde\Delta_k\equiv\tilde\Delta_k(i\tw)$. Except $\bar l_n(0)$, the threshold functions are $k$ dependent. However, if we approximate $\tDelta_k(i\tw)$ by its low-frequency behavior $\tw^2$ in the dimensionless propagator $\tilde G_k(\tilde q,i\tw)$, they all become $k$ independent. This approximation is sufficient to understand the ground state of the system and most of its physical properties, but an accurate determination of $\Delta_k(i\w)$ requires to keep the full frequency dependence of $\tDelta_k(i\tw)$~\cite{Dupuis20sm}. 

\begin{figure}
	\centerline{\includegraphics[width=15cm]{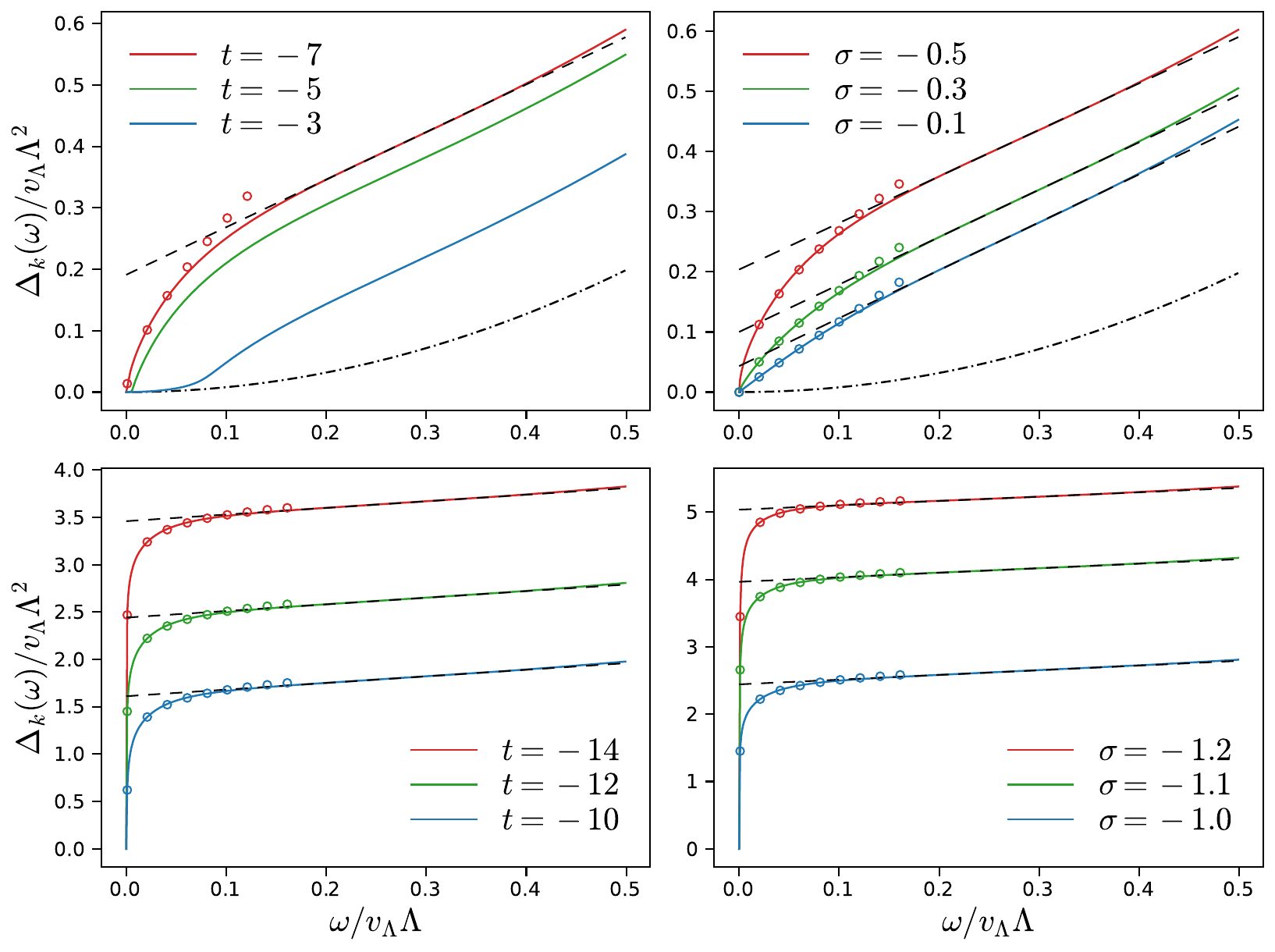}}
	\caption{Low-frequency behavior of the self-energy. The left figures show $\Delta_k(i\w)$ for various values of $t=\ln(k/\Lambda)$ ($\sig=-0.5$ (top) and $-1$ (bottom)) and the right ones correspond to various values of $\sig$ (with $t=-12$). The top figures are obtained for $x=2-\theta+3\sig/2>0$ and the bottom ones for $x<0$. In the last case, the self energy does not converge for $k\to 0$ but takes larger and larger values (see text). The dash-dotted line in the top figures shows the initial condition $\Delta_\Lambda(i\w)\propto\w^2$, the circles correspond to $A'_k+B'|\w|^{x/z}$ and the dashed lines to $\Delta_k(i\w)=A_k + B|\w|$.}
	\label{fig_self} 
\end{figure}

The relevance of disorder in the Wigner crystal phase can be determined from the linearized equation $\dt\delta_k=-(3+2\sig_k)\delta_k-K_kl_1 \delta_k''$ where $K_k\sim f_k^{-1/2}$ vanishes for $k\to 0$. The backward scattering is therefore a relevant perturbation for $\sig>-3/2$, and an irrelevant one for $\sig<-3/2$, in agreement with the scaling analysis discussed in Sec.~\ref{subsec_replica}. Solving numerically the full flow equations satisfied by $\delta_k(u)$ and $\tDelta_k(i\tw)$ in the case $-3/2<\sig\leq 0$, we find results that are reminiscent of the Bose-glass phase. The Luttinger parameter $K_k\sim k^{-\sig/2+\theta}$ vanishes with the exponent $\pm \sig/2+\theta$, whereas the velocity behaves as $v_k\sim k^{\theta+\sig/2}$. Thus the velocity vanishes in the limit $k\to 0$ if $\sig>-2\theta$ but diverges (as in the Wigner crystal) if $\sig<-2\theta$. Whether the latter case actually occurs (which requires $\theta<3/4$ since $\sig>-3/2$ in the Mott glass) depends on the value of $\theta$ which, for reasons explained in Ref.~\cite{Dupuis20sm}, cannot be accurately determined from the flow equations~(\ref{rgeq}) (in practice the value of $\theta$ depends on the choice of the regulator function $R_k$). Note that in any case the charge stiffness $D_k=v_kK_k\sim k^{2\theta}$ vanishes for $k\to 0$ since $\theta>0$ when $\sig>-3/2$. The disorder correlator reaches a fixed-point solution $\delta^*(u)$ with a $\sig$-dependent prefactor that coincides with the Bose-glass result when $\sig=0$ and vanishes when $\sig\to -3/2$ (see Eq.~(13) in the main text). 

As for the self-energy $\Delta_k(i\w)$, one can distinguish three regimes in the small $k$ limit (see Fig.~\ref{fig_self}). For $|\w|\lesssim v_kk$, $\Delta_k(i\w)\simeq Z_x\w^2/v_k^2$ (the corresponding frequency range is too narrow to be seen in Fig.~\ref{fig_self}). At higher frequencies, it behaves first as $(B'/x)|\w|^{x/z}+A'_k$ and then as $B|\w|+A_k$ up to the pinning frequency $\w_p\sim v(q_c)q_c$ determined by the Larkin length $L_c=q_c^{-1}\sim(Z_x^2/\calD)^{1/(3+2\sig)}$, where $x=2-\theta+3\sig/2$ and $z=\lim_{k\to 0}z_k$ is the dynamical exponent. When $\sig$ varies between 0 and $-3/2$, $x$ necessary changes from positive to negative since $\theta>0$. When $x>0$, i.e., for sufficiently small $|\sig|$, $A_k$ reaches a finite limit $A$ for $k\to 0$, as in the Bose-glass phase; in that case we expect that the self-energy converges nonuniformly toward the singular solution $\Delta_{k=0}(i\w)= (1-\delta_{\w,0}) (A+B|\w|)$ even though the truncation~(\ref{truncation}) does not allow to confirm this behavior at very low frequencies in the limit $k\to 0$~\cite{Dupuis20sm}. Such a self-energy implies that the conductivity is gapless, with a real part vanishing as $B\w^2/A^2$. On the other hand, when $x<0$, $A_k$ does not reach a finite limit when $k\to 0$ but seems to diverge. This could indicate that the conductivity vanishes with an exponent larger than 2: $\Re[\sig(\w)]\ll \w^2$.

\subsection{C.\;\; Density-density correlation function} 

The flow equations discussed so far are not sufficient to compute the disorder-averaged order parameter $\overline{\mean{e^{2i\varphi(x,\tau)}}}$ associated with the Wigner crystallization and the corresponding correlation function. Introducing a complex external source in the action, i.e., $S[\varphi;\xi]\to S[\varphi;\xi]+S_h[\varphi;h^*,h]$ with
\beq 
S_h[\varphi;h^*,h] = - \int_{x,\tau} [ h^*(x,\tau) e^{2i\varphi(x,\tau)} + \cc ]  ,
\eeq 
we have 
\beq
\begin{split}
	\overline{\mean{e^{2i\varphi(x,\tau)}}} &= \frac{\delta W_1[J,h^*,h]}{\delta h^*(x,\tau)} \bigg|_{J=h^*=h=0} , \\
	\chi(x,\tau;x',\tau') &= \overline{\mean{e^{2i\varphi(x,\tau)} e^{-2i\varphi(x',\tau')}}} 
	- \overline{\mean{e^{2i\varphi(x,\tau)}} \mean{e^{-2i\varphi(x',\tau')}} } = 
	\frac{\delta^2 W_1[J,h^*,h]}{\delta h^*(x,\tau) \delta h(x',\tau')} \bigg|_{J=h^*=h=0} ,
\end{split}
\label{eqsm1}
\eeq 
where $W_1[J,h^*,h]=\overline{W[J,h^*,h;\xi]}$ is the first cumulant of the random functional $W[J,h^*,h;\xi]=\ln \calZ[J,h^*,h;\xi]$. Equations~(\ref{eqsm1}) can be rewritten in terms of the Legendre transform $\Gamma_1[\phi,h^*,h]$ of $W_1[J,h^*,h]$~\cite{notsm1}, 
\beq 
\begin{gathered}
\overline{\mean{e^{2i\varphi(x,\tau)}}} = - \Gamma_1^{(010)}(x,\tau;\bar\phi) , \\
\chi(q,i\w) = -\Gamma_1^{(011)}(q,i\w;\bar\phi) + \Gamma_1^{(110)}(-q,-i\w;\bar\phi) G(q,i\w) \Gamma_1^{(101)}(q,i\w;\bar\phi) , 
\end{gathered}
\label{eqsm2}
\eeq 
where $\bar\phi\equiv\bar\phi[h^*=0,h=0]$ and the order parameter $\bar\phi[h^*,h]$ is defined by 
\beq 
\frac{\delta\Gamma_1[\phi;h^*,h]}{\delta\phi(x,\tau)}\biggl|_{\phi=\bar\phi[h^*,h]} = 0 
\eeq 
(we can assume $\bar\phi=0$ with no loss of generality). We use the notations 
\beq 
\begin{split}
	\Gamma_1^{(010)}[x,\tau;\phi,h^*,h] &= \frac{\delta \Gamma_1[\phi,h^*,h]}{\delta h^*(x,\tau)} , \\ 
	\Gamma_1^{(001)}[x,\tau;\phi,h^*,h] &= \frac{\delta \Gamma_1[\phi,h^*,h]}{\delta h(x,\tau)} , \\
	\Gamma_1^{(110)}[x,\tau;x',\tau';\phi,h^*,h] &= \frac{\delta^2 \Gamma_1[\phi,h^*,h]}{\delta\phi(x,\tau)\delta h^*(x',\tau')} ,
\end{split}
\eeq
etc., and denote by $\Gamma_1^{(nml)}(\cdots;\phi)$ the vertices evaluated in a constant field $\phi$ and $h^*=h=0$. $G=(\Gamma_1^{(200)})^{-1}$ denotes the disorder-averaged propagator for $h^*=h=0$ and is independent of the constant field $\phi$. 

When the regulator term $\Delta S_k$ is included in the action, the vertices $\Gamma_{1,k}^{(nml)}$ become $k$ dependent with initial values 
\beq
\Gamma^{(010)}_{1,\Lambda}[x,\tau;\phi,h^*,h] = -e^{2i\phi(x,\tau)} , \quad
\Gamma^{(001)}_{1,\Lambda}[x,\tau;\phi,h^*,h] = -e^{-2i\phi(x,\tau)} , \quad
\Gamma^{(011)}_{1,\Lambda}[x,\tau;x',\tau';\phi,h^*,h] = 0 .
\eeq
Using the fact that $\Gamma_{1,k}^{(211)}(x,\tau;x',\tau';\phi)$ remains equal to zero in the flow, we obtain
\beq
\begin{split}
	\dt \Gamma_{1,k}^{(010)}(x,\tau) &= \half \tdt \tr \llbrace G_k \Gamma_{1,k}^{(210)}(x,\tau) \left[1 + G_k \Gamma^{(11)}_{2,k} \right] \rrbrace , \\ 
	\dt \Gamma_{1,k}^{(001)}(x,\tau) &= \half \tdt \tr \llbrace G_k \Gamma_{1,k}^{(201)}(x,\tau) \left[1 + G_k \Gamma^{(11)}_{2,k} \right] \rrbrace 
\end{split}
\eeq 
and 
\begin{align}
\dt \Gamma_{1,k}^{(011)}(x,\tau;x',\tau') ={}& - \half \tdt \tr \Bigl\{  G_k \Gamma_{1,k}^{(201)}(x',\tau') G_k \Gamma_{1,k}^{(210)}(x,\tau) \nonumber\\ & 
+ G_k \Gamma_{1,k}^{(201)}(x',\tau') G_k \Gamma_{1,k}^{(210)}(x,\tau) G_k \Gamma_{2,k}^{(11)}  
+ G_k \Gamma_{1,k}^{(210)}(x,\tau) G_k \Gamma_{1,k}^{(201)}(x',\tau') G_k \Gamma_{2,k}^{(11)} 
\Bigr\} ,
\end{align}
where the trace is over space and time (or momentun and frequency) variables and $\tdt=(\dt R_k)\partial_{R_k}$ acts only on the $t$ dependence of $R_k$. To alleviate the notations, we do not write explicitly the dependence on $\phi$. The solution is of the form $\Gamma_{1,k}^{(010)}(x,\tau;\phi)=-A_k e^{2i\phi}$ and $\Gamma_{1,k}^{(001)}(x,\tau;\phi)=-A_k e^{-2i\phi}$, with $A_\Lambda=1$, so that $\overline{\mean{e^{2i\varphi(x,\tau)}}}=\lim_{k\to 0} A_k$. We thus obtain
\beq 
\begin{split}
	\dt \ln A_k &= 2 K_k l_1 + 2 \pi \delta_k(0) \bar l_2 , \\
	\dt \Gamma_{1,k}^{(011)}(q=0,i\w=0) &= \frac{16\pi A_k^2 K_k}{k^2 v_k} \left( \frac{K_k}{2} l_2 + \pi \delta_k(0) \bar l_3 \right)  
\end{split}
\label{eqsm3}
\eeq
and 
\beq 
\chi_k(q=0,i\w=0) = - \Gamma_{1,k}^{(011)}(q=0,i\w=0) - 4 A_k^2 G_k(q=0,i\w=0) . 
\label{eqsm4}
\eeq
Solving these equations, we obtain $\chi_k(q=0,i\w=0)\sim k^\alpha$ and therefore the scaling dimension $[\chi(q,i\w)]=\alpha$. By dimensional analysis, we then deduce $[\chi(x,\tau)]=1+z+[\chi(q,i\w)]=1+z+\alpha$ and in turn $\chi(x)\equiv \chi(x,\tau=0)~\sim 1/|x|^{1+z+\alpha}$ where $z$ is the dynamical critical exponent. 

\subsubsection{1.\;\; Luttinger liquid} 

Let us show that we recover the expected long-distance behavior of $\chi(x)$ in a Luttinger liquid. Solving~(\ref{eqsm3}) with $K_k=\KLL$, $v_k=\vLL$ and $\delta_k(0)=0$, we obtain $A_k=k^{\KLL}$ using $l_1=1/2$~\cite{notsm3}. The vanishing of $A_k$ when $k\to 0$ implies the absence of Wigner crystallization. Using the expression of $A_k$ and $G_k(q=0,i\w=0)=1/R_k(q=0,i\w=0)\sim k^{-2}$, we find that both terms in the rhs of~(\ref{eqsm4}) vary as $k^{2\KLL-2}$, which gives $\alpha=2\KLL-2$ and $\chi(x)\sim 1/|x|^{2\KLL}$ using $z=1$. 

\subsubsection{2.\;\; Wigner crystal} 

When $K_k\sim k^{-\sig/2}$ and $v_k\sim k^{\sig/2}$ (with $\sig<0$, we do not consider the case of Coulomb interactions), $A_k$ reaches a nonzero limit for $k\to 0$, signaling Wigner crystallization, and $\Gamma_{1,k}^{(011)}(0)\sim k^{-2-3\sig/2}$. The second term in the rhs of~(\ref{eqsm4}) varies as $A_k^2G_k(0)\sim 1/R_k(0)\sim k^{-2-\sig}$ and therefore is the dominant contribution to $\chi(q=0,i\w=0)$. This gives $\alpha=-2-\sig$ so that $\chi(x)\sim 1/|x|^{-\sig/2}$, using $z=1+\sig/2$, in agreement with the direct calculation from~(\ref{eqsm12}).

\subsubsection{3.\;\; Bose glass and Mott glass}

Since $K_k\sim k^{\theta-\sig/2}\to 0$ in the Bose ($\sig=0$) or Mott ($\sig<0$, we do not consider the case of Coulomb interactions) glass, Eqs.~(\ref{eqsm3}) simplify into 
\beq 
\begin{split}
	\dt \ln A_k &= 2 \pi \delta^*(0) \bar l_2 , \\
	\dt \Gamma_{1,k}^{(011)}(q=0,i\w=0) &= \frac{16\pi^2 A_k^2 K_k}{k^2 v_k} \delta^*(0) \bar l_3  
\end{split}
\label{eqsm5}
\eeq
when $k\to 0$. This gives $A_k\sim k^{2\pi \bar l_2\delta^*(0)}\sim k^{\pi^2(3+2\sig)/18}$ and $\Gamma_{1,k}^{(011)}(0)\sim k^{\pi^2(3+2\sig)/9-2-\sig}$. The second term in the rhs of~(\ref{eqsm4}) yields a similar contribution so that $\alpha=\pi^2(3+2\sig)/9-2-\sig$, using $z=1+\sig/2+\theta$, and in turn $\chi(x)\sim 1/|x|^{\pi^2(3+2\sig)/9+\theta-\sig/2}$. 

\input{SM_final.bbl}


\end{document}

%% file: definition.tex

\def\rhoeq{\hat\rho_{\rm eq}}

\newcommand{\marge}[1]{\marginpar{\scriptsize #1}}
\newcommand{\remarque}[1]{\marginpar{\scriptsize Remarque}{\it [#1]}}
\newcommand{\new}[1]{{\bf #1}}
\newcommand{\red}[1]{\textcolor{red}{#1}}
\newlength{\textlarg}
\newcommand{\redbar}[1]{\textcolor{red}{\st{#1}}} 
\newcommand{\bluebar}[1]{\textcolor{blue}{\st{#1}}} 

\newcommand{\beq}{\begin{equation}}
\newcommand{\eeq}{\end{equation}}
\newcommand{\bfig}{\begin{figure}}
\newcommand{\efig}{\end{figure}}
\newcommand{\bline}{\begin{multline}}
\newcommand{\eline}{\end{multline}}
\newcommand{\bremark}{\begin{quotation} \noindent \small }
\newcommand{\eremark}{\end{quotation}}
\newcommand{\llbrace}{\left\lbrace}  
\newcommand{\rrbrace}{\right\rbrace}
\newcommand{\lbraket}{\left[}
\newcommand{\rbraket}{\right]}
\newcommand{\llangle}{\left\langle}
\newcommand{\rrangle}{\right\rangle} 

\newcommand{\Tr}{{\rm Tr}} 
\newcommand{\tr}{{\rm tr}} 
\newcommand{\sgn}{\,{\rm sgn}} 
\newcommand{\mean}[1]{\langle #1 \rangle}
\newcommand{\commu}[2]{[#1,#2]} 
\newcommand{\bra}[1]{\langle#1|}
\newcommand{\ket}[1]{|#1\rangle}
\newcommand{\braket}[2]{\langle #1|#2\rangle}
\newcommand{\ketbra}[2]{|#1\rangle\langle#2|}
\newcommand{\dbraket}[3]{\langle #1|#2|#3\rangle}
\newcommand{\tens}[1]{\overleftrightarrow{#1}}  
\newcommand{\vac}{|{\rm vac}\rangle} 
\newcommand{\bravac}{\langle{\rm vac}|}
\newcommand{\const}{{\rm const}} 
\newcommand{\unif}{{\rm unif.}} 
\newcommand{\atanh}{\,{\rm atanh}}
\newcommand{\cotanh}{\,{\rm cotanh}}

\newcommand{\ie}{i.e.\xspace}
\newcommand{\iet}{i.e.}
\newcommand{\eg}{e.g.\xspace}
\newcommand{\cc}{{\rm c.c.}} 
\newcommand{\hc}{{\rm h.c.}} 
\newcommand{\etal}{{\it et al. }}
\newcommand\eme{$^{\mbox{\footnotesize ème}}$\xspace}

\newcommand{\jhatbf}{\hat {\textbf \jold}} 
\newcommand{\Jhatbf}{\hat {\textbf \J}} 
\newcommand{\jhat}{\hat {\jmath}} 
\newcommand{\Jhat}{\hat {J}} 
\newcommand{\jbf}{\textbf j}
\newcommand{\Jbf}{\textbf J}

\def\chibf{\boldsymbol{\chi}}
\def\down{\downarrow}
\def\eps{\epsilon}
\def\gam{\gamma} 
\def\alphabf{\boldsymbol{\alpha}}
\def\phibf{\boldsymbol{\phi}}
\def\varphibf{\boldsymbol{\varphi}}
\def\varphibfs{\boldsymbol{\varphi}_<}
\def\varphibfl{\boldsymbol{\varphi}_>}
\def\varphis{\varphi_{<}}
\def\varphil{\varphi_{>}}
\def\psibf{\boldsymbol{\psi}}
\def\thetabf{\boldsymbol{\theta}}
\def\Ome{\Omega}
\def\omeD{{\omega_D}} 
\def\bfOme{\boldsymbol{\Omega}} 
\def\Omebf{\boldsymbol{\Omega}} 
\def\lamb{\lambda}
\def\Lamb{\Lambda}
\def\sig{\sigma}
\def\Sig{\Sigma}
\def\sigp{{\sigma'}} 
\def\bfsig{\boldsymbol{\sigma}} 
\def\sigbf{\boldsymbol{\sigma}} 
\def\bfSig{\boldsymbol{\Sigma}} 
\def\The{\Theta} 
\def\up{\uparrow}

\def\epsk{\epsilon_{\bf k}} 
\def\xik{\xi_{\bf k}} 
\def\txik{\tilde\xi_{\bf k}} 
\def\xip{\xi_{\bf p}} 
\def\xiq{\xi_{\bf q}} 
\def\xikq{\xi_{{\bf k}+{\bf q}}} 
\def\Ek{E_{\bf k}} 
\def\Ep{E_{\bf p}}
\def\Eq{E_{\bf q}}
\def\Heff{\hat H_{\rm eff}}
\def\Hem{\hat H_{\rm em}}
\def\Hint{\hat H_{\rm int}}
\def\Hloc{\hat H_{\rm loc}}
\def\HMF{\hat H_{\rm MF}}
\def\HLL{\hat H_{\rm LL}}
\def\Sem{S_{\rm em}}
\def\SMF{S_{\rm MF}} 
\def\SHF{S_{\rm HF}} 
\def\SRPA{S_{\rm RPA}} 
\def\Sint{S_{\rm int}} 
\def\Sloc{S_{\rm loc}}
\def\TN{T_{\rm N}} 
\def\TNHF{T^{\rm HF}_{\rm N}} 
\def\Zloc{Z_{\rm loc}} 
\def\ZMF{Z_{\rm MF}} 
\def\ZHF{Z_{\rm HF}} 
\def\ZRPA{Z_{\rm RPA}} 
\def\RPA{{\rm RPA}}
\def\loc{{\rm loc}} 
\def\pp{{\rm pp}}
\def\ph{{\rm ph}} 
\def\ch{{\rm ch}}
\def\sp{{\rm sp}} 
\def\qtf{q_{\rm TF}}
\def\epstf{\eps^{}_{\rm TF}} 
\def\epsrpa{\eps^{}_{\rm RPA}} 
\def\chinnzpp{\chi_{nn}^{0}{}\!\!\!''}

\def\half{\frac{1}{2}}
\def\dhalf{\dfrac{1}{2}}
\def\third{\frac{1}{3}} 
\def\quarter{\frac{1}{4}}

\def\qr{{\bf q}\cdot{\bf r}}
\def\wt{\omega t} 

\def\a{{\bf a}}
\def\b{{\bf b}}
\newcommand{\cv}{{\bf c}} 
\def\e{{\bf e}}
\def\f{{\bf f}}
\def\g{{\bf g}}
\def\h{{\bf h}}
\def\jold{\char"11}
\def\j{{\bf j}}
\def\k{{\bf k}}
\def\l{{\bf l}}
\def\m{{\bf m}}
\def\n{{\bf n}} 
\def\p{{\bf p}} 
\def\q{{\bf q}}
\def\r{{\bf r}}
\def\t{{\bf t}}
\def\u{{\bf u}}
\newcommand{\vv}{{\bf v}}
\def\x{{\bf x}}
\def\y{{\bf y}} 
\def\z{{\bf z}} 
\def\A{{\bf A}}
\def\B{{\bf B}}
\def\D{{\bf D}} 
\def\E{{\bf E}} 
\def\F{{\bf F}} 
\def\H{{\bf H}}  
\def\J{{\bf J}}
\def\K{{\bf K}} 

\def\G{{\bf G}}
\def\L{{\bf L}}
\def\M{{\bf M}}  
\def\O{{\bf O}} 
\def\P{{\bf P}} 
\def\Q{{\bf Q}} 
\def\R{{\bf R}}
\def\S{{\bf S}}
\def\U{{\bf U}} 
\def\V{{\bf V}} 
\def\X{{\bf X}} 
\def\Y{{\bf Y}} 
\def\epsbf{\boldsymbol{\epsilon}}
\def\betabf{\boldsymbol{\beta}}
\def\deltabf{\boldsymbol{\delta}}
\def\mubf{\boldsymbol{\mu}}
\def\nablabf{\boldsymbol{\nabla}}
\def\rhobf{\boldsymbol{\rho}}
\def\sigmabf{\boldsymbol{\sigma}} 
\def\Pibf{\boldsymbol{\Pi}}
\def\pibf{\boldsymbol{\pi}}

\def\para{\parallel}
\def\kpara{{k_\parallel}}
\def\kperp{{k_\perp}} 
\def\kperpp{{k_\perp'}} 
\def\qperp{{q_\perp}} 
\def\tperp{{t_\perp}} 

\def\w{\omega}
\def\wn{\omega_n}
\def\wm{\omega_m}
\def\wnu{\omega_\nu}
\def\wp{\omega_p} 
\def\dmu{{\partial_\mu}}
\def\dnu{{\partial_\nu}}
\def\dl{{\partial_l}}  
\def\dt{\partial_t} 
\def\tdt{\tilde\partial_t}
\def\dk{\partial_k}
\def\tdk{\tilde\partial_k}
\def\dx{\partial_x}
\def\dy{\partial_y} 
\def\dtau{{\partial_\tau}}  
\def\det{{\rm det}} 
\def\Pf{{\rm Pf}}
\def\diag{{\rm diag}}

\def\dsum{\displaystyle \sum}
\def\dint{\displaystyle \int} 
\def\intt{\int_{-\infty}^\infty dt} 
\def\inttp{\int_{-\infty}^\infty dt'} 
\def\intk{\int_{\bf k}} 
\def\intkd{\int \frac{d^dk}{(2\pi)^d}}
\def\intq{\int_{\bf q}} 
\def\intr{\int d^dr}  
\def\dintr{\displaystyle \int d^dr} 
\def\intrp{\int d^dr'}
\def\dinttau{\displaystyle \int_0^\beta d\tau}
\def\dinttaup{\displaystyle \int_0^\beta d\tau'}
\def\inttau{\int_0^\beta d\tau}
\def\inttaup{\int_0^\beta d\tau'}
\def\intx{\int d^{d+1}x} 
\def\inttaur{\int_0^\beta d\tau \int d^dr}
\def\intinf{\int_{-\infty}^\infty}
\def\dinttaur{\displaystyle \int_0^\beta d\tau \int d^dr}
\def\dintinf{\displaystyle \int_{-\infty}^\infty}
\def\intw{\int_{-\infty}^\infty \frac{d\w}{2\pi}}
\def\sumr{\sum_{\bf r}} 

\def\calA{{\cal A}}
\def\calAbf{\bm{{\cal A}}}
\def\calB{{\cal B}} 
\def\calC{{\cal C}} 
\def\dt{\partial_t}
\def\calD{{\cal D}}
\def\calE{{\cal E}}
\def\calF{{\cal F}} 
\def\calFbf{\bm{{\cal F}}}
\def\calG{{\cal G}}
\def\calH{{\cal H}}
\def\calI{{\cal I}}
\def\calJ{{\cal J}}
\def\calK{{\cal K}}
\def\calL{{\cal L}} 
\def\calM{{\cal M}} 
\def\calN{{\cal N}}
\def\calO{{\cal O}}
\def\calP{{\cal P}}  
\def\calR{{\cal R}} 
\def\calS{{\cal S}}
\def\calT{{\cal T}}
\def\calU{{\cal U}}
\def\calV{{\cal V}}
\def\calX{{\cal X}} 
\def\calY{{\cal Y}} 
\def\calZ{{\cal Z}} 

\def\calbfB{{\bf \cal B}}
\def\calbfF{{\bf \cal F}}

\def\tT{{\tilde T}}
\def\talpha{{\tilde\alpha}}
\def\tbeta{{\tilde\beta}}
\def\tchi{{\tilde\chi}}
\def\tdelta{{\tilde\delta}}
\def\tDelta{{\tilde\Delta}}
\def\teta{{\tilde\eta}} 
\def\tlamb{{\tilde\lambda}}
\def\tmu{{\tilde\mu}}
\def\tphibf{{\tilde\phibf}}
\def\trho{{\tilde\rho}}
\def\tvarphibf{{\tilde\varphibf}} 
\def\tw{{\tilde\omega}}
\def\twn{{\tilde\omega_n}}
\def\twnu{{\tilde\omega_\nu}}

\def\asinh{{\rm asinh}} 
\def\Tbkt{T_{\rm BKT}}

%% file: MG_final.bbl
%

%% file: SM_final.bbl
%

%% file: MGarXiv.bbl
\begin{thebibliography}{47}
\bibitem [{not({\natexlab{a}})}]{not4}%
  \BibitemOpen
  \href@noop {} {} \bibinfo {note} {The short-distance
  cutoff $a$ is necessary to make the Fourier transform of $\calV_\sig(x)$ well
  defined when $\sig\geq 0$.}\BibitemShut {Stop}%
\bibitem [{not({\natexlab{b}})}]{not3}%
  \BibitemOpen
  \href@noop {} {} \bibinfo {note} {At zero temperature, the
  charge stiffness $D$ is related to the superfluid density
  $\rho_s=D/\pi$.}\BibitemShut {Stop}%
\bibitem [{not({\natexlab{d}})}]{not5}%
  \BibitemOpen
  \href@noop {} {} \bibinfo {note} {For sufficiently small
  disorder, one always has $L_c>L_x$.}\BibitemShut {Stop}%
\bibitem [{not({\natexlab{f}})}]{not6}%
  \BibitemOpen
  \href@noop {} {} \bibinfo {note} {This result differs from
  the low-frequency conductivity $\sig(\w)\sim |\w|$ (ignoring logarithmic
  corrections) obtained by Shklovskii and Efros for the electron gas with
  Coulomb interactions~\cite{Shklovskii81}. This expression was however
  obtained in the strong-disorder limit where the localization length is much
  smaller than the interparticle distance whereas the bosonization approach is
  valid in the opposite, weak-disorder, limit~\cite{Maurey95}.}\BibitemShut
  {Stop}%
\end{thebibliography}
\begin{thebibliography}{24}
\bibitem [{not({\natexlab{a}})}]{notsm2}%
  \BibitemOpen
  \href@noop {} {} \bibinfo {note}
  {$[e^{2i\varphi(x,\tau)}]=\KLL$ in the Luttinger liquid. In the Wigner
  crystal phase, since $K_k\to 0$, we can assign a vanishing scaling dimension
  to $e^{2i\varphi(x,\tau)}$.}\BibitemShut {Stop}%
\bibitem [{not({\natexlab{c}})}]{notsm3}%
  \BibitemOpen
  \href@noop {} {} \bibinfo {note} {The value $l_1=1/2$ in
  the Luttinger-liquid phase is universal, i.e., independent of the function
  $r(x)$.}\BibitemShut {Stop}%
\end{thebibliography}
